\documentclass[aps,prl,twocolumn,superscriptaddress,groupedaddress]{revtex4}
\usepackage[centertags]{amsmath}
\usepackage{amsfonts}
\usepackage{amssymb}
\usepackage{amsthm}
\usepackage{newlfont}
\usepackage{stmaryrd}
\usepackage{mathrsfs}
\usepackage{euscript}
\usepackage{graphicx}
\usepackage{enumerate}

\usepackage{wrapfig}
\usepackage{subfigure}
\usepackage{amscd}
\usepackage{hyperref}

\theoremstyle{plain}

\theoremstyle{definition}

\theoremstyle{remark}

\let\al=\alpha \let\be=\beta \let\de=\delta 
\let\ve=\varepsilon  \let\ga=\gamma 
 \let\la=\lambda  
   
\let\ze=\zeta 

 \let\Ga=\Gamma \let\La=\Lambda \let\Om=\Omega

\newcommand{\caI}{{\mathcal I}}
\newcommand{\caJ}{{\mathcal J}}

\newcommand{\caX}{{\mathcal X}}

\newcommand{\caZ}{{\mathcal Z}}

\newcommand{\bbN}{{\mathbb N}}

\newcommand{\opunit}{\text{1}\kern-0.22em\text{l}}

\newcommand{\bsP}{{\boldsymbol P}}

\DeclareMathAlphabet{\mathpzc}{OT1}{pzc}{m}{it}

\newcommand{\id}{\textrm{d}}

\begin{document}

\title{Non-reactive forces and pattern formation\\
induced by a nonequilibrium medium}

\author{Christian Maes}
\affiliation{Instituut voor Theoretische Fysica, KU Leuven, Belgium}
\author{Karel Neto\v{c}n\'{y}}
\email{netocny@fzu.cz}
\affiliation{Institute of Physics, Academy of Sciences of the Czech Republic, Prague, Czech Republic}

\begin{abstract}
We study the induced interaction between multiple probes locally interacting with driven colloids and trapped in a toroidal geometry. The effective binary forces between the probes break the action-reaction principle and their range decreases with the driving.  We demonstrate how in the stationary nonlinear nonequilibrium regime these interactions induce stability of a crystal-like pattern, where the probes are equidistant, when the probe-colloid interaction is either completely attractive or completely repulsive.
\end{abstract}

\maketitle

\section{Introduction}

Self-assembly and pattern formation are important subjects of contemporary research both for its technological and fundamental significance~\cite{selfassem}.  While one can try to exploit and even manufacture specific interactions between constituents possibly combined with special boundary conditions, it is clear that nature itself utilizes more robust mechanisms in giving a large variety of beautiful spatio-temporal patterns~\cite{morphog}.  It is generally believed that nonequilibrium
conditions are often at the origin of such structures, but there are few
first-principle studies of the emergence of such patterns from contact with nonequilibrium media.
Indeed one would wish to avoid specially constructed or {\it ad hoc} modeling to deduce such structures.
The present Letter provides possibly the simplest such scenario where calculations can be done
explicitly and the emergence of a stationary spatial pattern can be followed in great detail.  The mesoscopic model is simple and, we believe,
feasible to reproduce in the laboratory.  The microscopic interaction and details are largely irrelevant for the reported phenomena which at the same time leads to a natural robustness of the induced pattern.  
either completely repulsive or completely attractive.  
The emerging stable pattern is crystal-like as a result of
the effective repulsive forces between the probes.
Yet the induced interaction, while via binary forces, is not satisfying the action-reaction law, e.g. like in~\cite{cas,cas1}.
Moreover, though the magnitude of the induced force readily increases with driving, its range decreases and the force becomes more local. The stability is then an intriguing effect which is optimally enhanced for intermediate driving magnitude.

Induced stability from contact with a nonequilibrium medium has been discussed before in e.g.~\cite{njp,Sheshka, epl,act}, and very recently also for crystallization as in the present paper; see \cite{cry}.  The nature of statistical forces induced by nonequilibrium baths has e.g. been explored in~\cite{mal,yan,kaf,prl,tim}.  Our Letter treats stabilization of patterns among multiple probes as a topic in  nonequilibrium statistical mechanics, which is foundational for the understanding of morphogenesis and self-assembly outside equilibrium.

\begin{figure}[t]
\includegraphics[width=3cm]{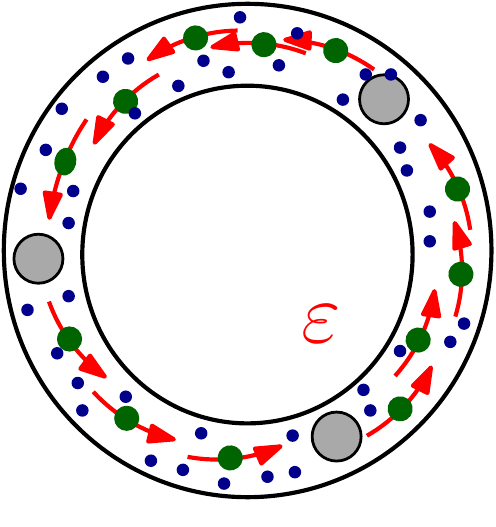}\hspace*{0.2cm}
\includegraphics[width=4.2cm]{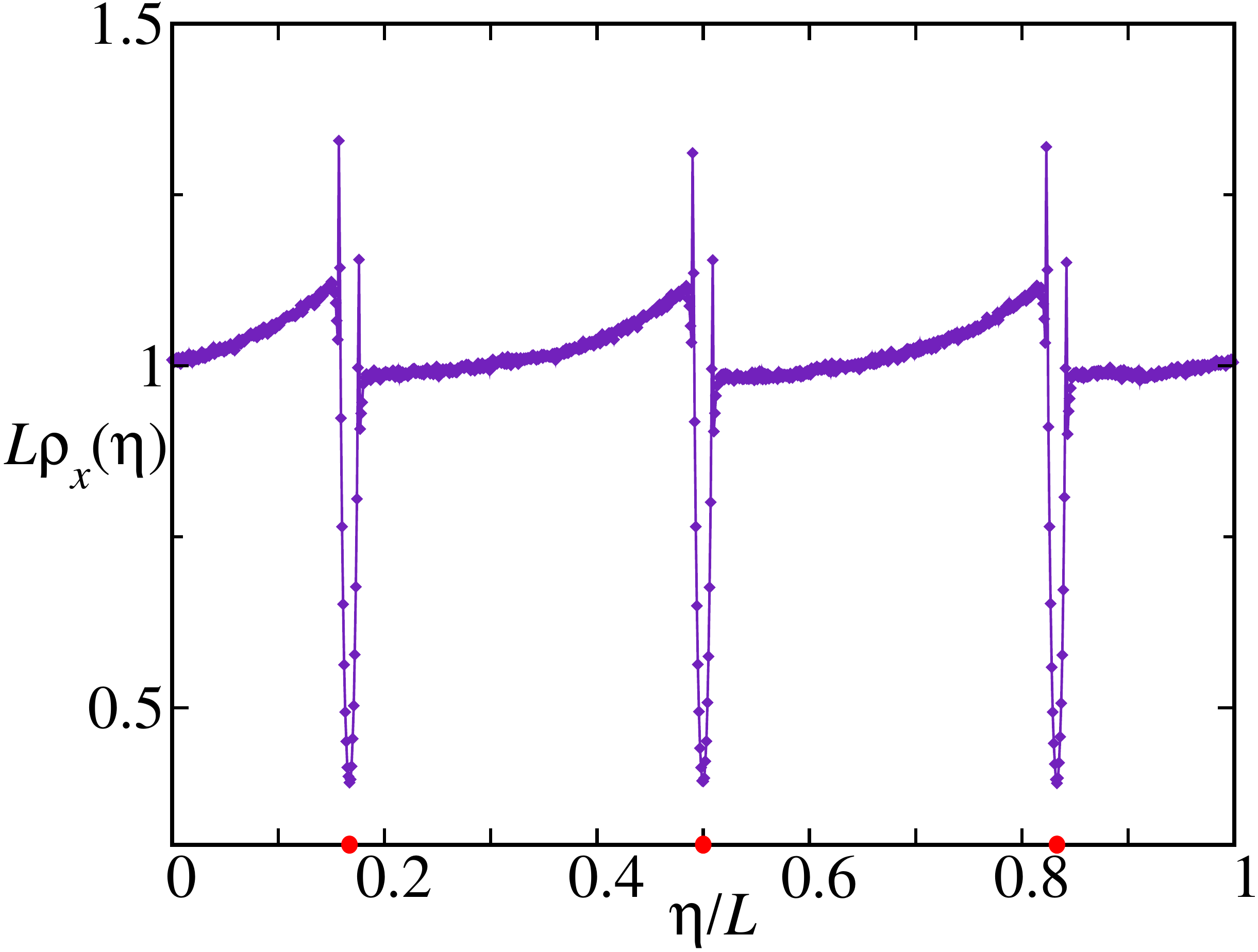}
\caption{\small{(Left) Passive probes suspended in a driven colloidal fluid.  The arrowed particles are the driven colloids (via rotational force $\varepsilon$) and the thermal environment is represented by the many smaller particles in the background. (Right) Density of colloids with interaction
$u(z) = [1- (z/\delta)^2]^2$ for $|z|\leq \delta= 0.1$ and $\ve=1,  \beta=1, L=10$, for three equidistant probes.}}
\label{ring}
\end{figure}

\vspace{2mm}
\emph{The model} --- Identical probes at positions $x = (x_\al)_{\al=1}^N$ are suspended in a colloidal fluid; see the cartoon in
Fig.~\ref{ring}(left).  The colloids are sufficiently dilute and are modeled here via independent particles with generic coordinate $\eta$ and subject to thermal noise.  All particles are trapped on a ring of length $L$. The number of colloids is $\rho^o L$ for some (fixed) average density $\rho^o$. The probes are thought to be heavy and moving on a much larger time-scale than the colloids which allows their treatment in the quasi-static limit. The probes and colloids are coupled via a (sufficiently smooth) local interaction potential
$u(x_\al - \eta)$ such that $u(-z) = u(z)$, and $u(z) = 0$ whenever $|z| > \de$; the parameter
$\de$ specifying the range of the interaction.
The dynamics of the colloids is driven as modelled by the overdamped diffusion,
\begin{equation}\label{smoluchowski}
\ze \frac{\id\eta_t}{\id t} = \ve - \frac{\partial U(x,\eta_t)}{\partial\eta} +
\Bigl( \frac{2\ze}{\be} \Bigr)^{1/2} \xi_t\,
\quad (\eta\,\, \text{mod } L)
\end{equation}
with $U(x,\eta) = \sum_\al u(x_\al - \eta)$ the total potential and $\ve \geq 0$ a constant driving force.  Inverse temperature $\be = 1/(k_B T)$ and friction $\ze$ are fixed parameters.
The mean force on the $\alpha$th probe from the driven fluid is given by 
\begin{equation}\label{forces}
f_\al(x) = -\int_0^L u' (x_\al - \eta)\,\rho_x(\eta)\,\id\eta
\end{equation}
where $\rho_x$ is the stationary density of the colloids satisfying the stationary Smoluchowski equation
\begin{eqnarray}\label{stat}
\ze j_x &=& \rho_x(\eta)\,\bigl[ \ve + \sum_\al u'(x_\al - \eta) \bigr] -
\frac{\rho'_x(\eta)}{\be}
\end{eqnarray}
under periodic boundary conditions and for fixed colloid mass
$\int_0^L \rho_x(\eta)\,\id\eta = \rho^o L$; the constant
$j_x$ is the stationary current of the colloid flow and it depends on the probe positions $x$.
The forces~\eqref{forces} satisfy the transmission rule
\begin{equation}\label{force-current}
\sum_\al f_\al(x) = (\ve \rho^o - \ze j_x) L
\end{equation}
by which the enhancement of the total force, e.g., via tuning the probe-colloid interaction
$u(z)$, always
goes along with suppression of the colloid flow.
Combining \eqref{force-current} with the general inequality
$0 \leq j_x \leq \ve\rho^o / \ze$, cf.~\cite{SM}, implies the bound
$0 \leq \sum_\al f_\al(x) \leq \ve \rho^o L$ for the total force.

The detailed structure of the forces $f_\al(x)$ of course depends on how the probes modify the stationary colloidal density. Under equilibrium conditions
($\ve = 0$) this modification is \emph{local}
(on the microscopic scale $\de$) and given by the Boltzmann factor,
$\rho^\text{eq}_x(\eta) \propto \rho^o \exp\,[-\be U(x,\eta)]$. In particular, for
\emph{isolated} probes, i.e., when $ |x_\al - x_\ga| \geq 2\de$ pairwise, then
the colloidal density has an almost homogeneous profile modified by $N$ identical ``blobs'' around the positions of the probes,
\begin{equation}\label{density-eq}
\rho_x^\text{eq}(\eta) \propto \rho^o\,
\bigl[ 1 + \sum_\al \phi^-(x_\al - \eta) \bigr]
\end{equation}
with the Mayer functions $\phi^-(z) = e^{-\be u(z)} - 1$. Since the blobs are symmetric and their supports do not intersect, under equilibrium all forces $f_\al^\text{eq}(x)=0$ vanish for isolated probes.
Still, non-vanishing forces within clusters of non-isolated probes exist as mediated by colloids in the neighborhood of multiple probes. Moreover, an additional interaction between the colloids would generate density-density correlations which again manifest themselves as effective multiprobe forces. The latter can be observed on distances below the colloidal correlation length and they always satisfy the action-reaction principle.
However, here we report on non-standard effects which are \emph{entirely induced by nonequilibrium driving} and are associated with a (driving-dependent) length scale. For clarity and simplicity we assume isolated probes and noninteracting colloids, so that the equilibrium forces strictly vanish. Note that the isolation condition imposes an upper bound on the averaged density of probes,
$N/L \leq (2\de)^{-1}$.

\begin{figure}[t]
  \includegraphics[width=8cm]{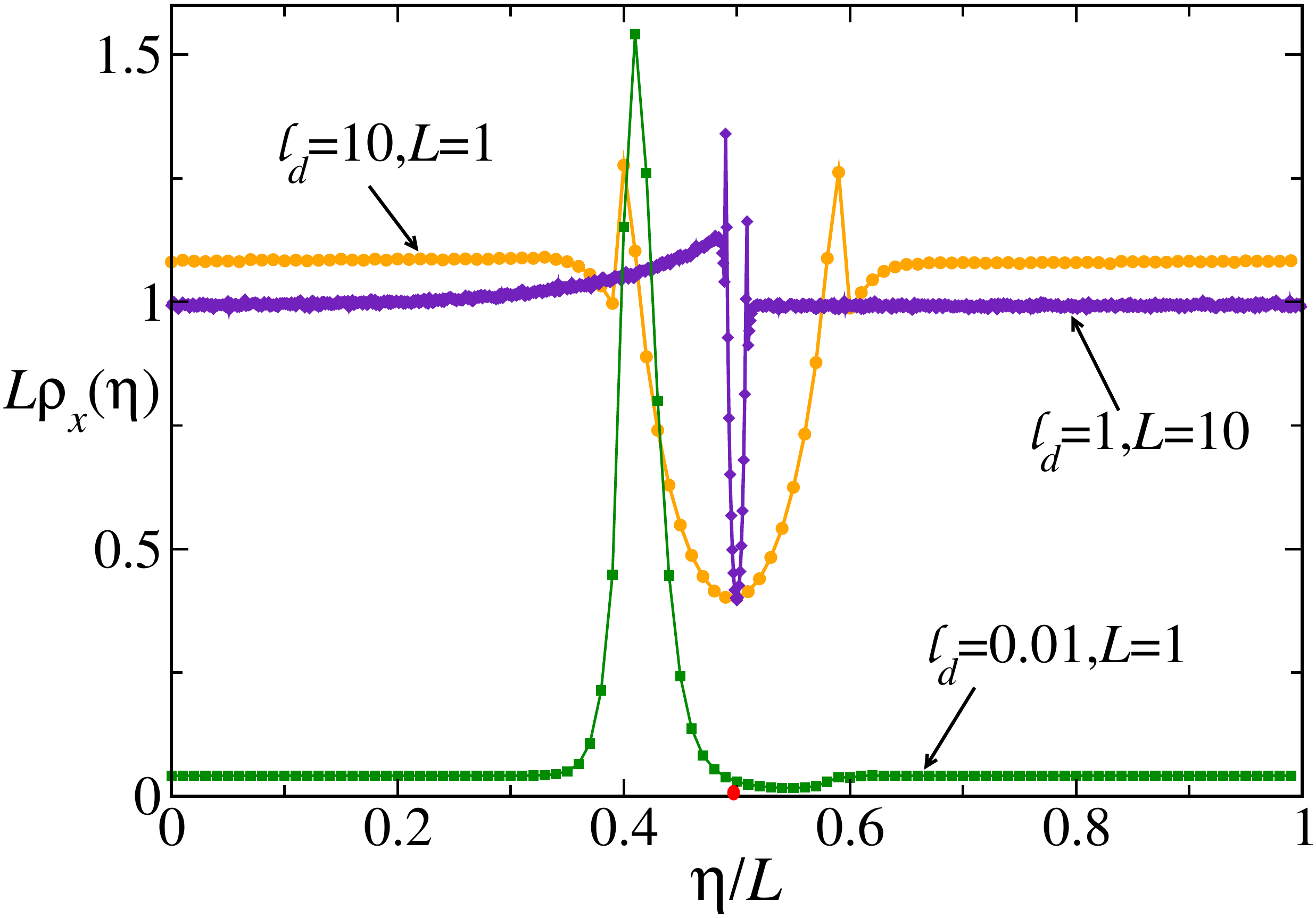}\\
  \caption{Density of colloids with interaction $u(z) = [1- (z/\delta)^2]^2$ for $|z|\leq \delta= 0.1$ around a single probe. Three regimes depending on relative length scales
  $\ell_d =(\ve \beta)^{-1}$, $L$ and $\delta$. The density for $\ell_d=0.01$ is scaled down by a factor of 10.}
\label{density-blob}
\end{figure}

\vspace{2mm}
\emph{Colloidal density tails} ---
For driven colloids \mbox{($\ve > 0$)} the mirror symmetry of the stationary density ``blobs'' around the probe positions
$x_\al$ breaks down; see Figs.~\ref{ring}(right)--\ref{density-blob}.
That relates to the emergence of a new length scale associated with the driving,
$\ell_d = (\be\ve)^{-1}$. The latter can be understood as a typical distance on which the dissipation as measured via the entropy flux to the thermal bath becomes relevant.  The interplay between the three length scales
$\de$, $\ell_d$ and $L$ plays a crucial role in the analysis.

The colloidal medium is (globally) close to equilibrium as long as the entropy flux for a colloid moving along the complete ring is small, $\be\ve L \ll 1$ (or $\ell_d \gg L$). Increasing the ring size $L$ we effectively drive the colloidal fluid further from equilibrium. In particular, in the (globally) strong nonequilibrium regime,
$\be\ve L \gg 1$ (or $L \gg \ell_d$), the finite size of the ring becomes largely irrelevant and one can pass to the thermodynamic limit. When also \emph{locally} --- on the length scale $\de$ --- the fluid is strongly driven, $\be\ve \de \gg 1$ (or $\ell_d \ll \de$), then the nonequilibrium density ``blobs'' again become local; cf.~Fig.~\ref{density-blob}. It is therefore natural to concentrate on the most interesting regime of globally strong ($\ell_d \ll L$) but locally not so strong ($\ell_d \gtrsim 2\de$) nonequilibrium.

From Fig.~\ref{density-blob} we already see that the driving-induced asymmetry in the colloidal density around each probe becomes fully developed in the (globally) strong driving regime
($L \gg \ell_d$).  Observe also that each probe creates a tail in the colloidal density which aims in the opposite direction with respect to the driving, whereas in the positive direction the density remains nearly unaffected. The tail has an exponential shape with $\ell_d = (\be\ve)^{-1}$ the characteristic length of decay.  That can be analytically derived from solving \eqref{stat} for the stationary density which obtains the asymptotic form, up to negligible corrections $O(e^{-L/\ell_d})$, cf.~\cite{SM},
\begin{equation}\label{density-neq}
\rho_x(\eta) = \frac{\rho^o e^{-\be U(x,\eta)}}{Z(x)}\,\Bigl[
1 - \frac{1}{\ell_d} \sum_\al \Phi(x_\al - \eta)\, e^{-(x_\al - \eta) / \ell_d} \Bigr]
\end{equation}
where $\Phi(z) = \int_{-\de}^z \id z'\,\phi^+(z')\,e^{z'/\ell_d}$, with
$\phi^+(z) = 1 - e^{\be u(z)}$, is a smooth interpolation between zero for
$z \leq -\de$ and a constant for $z \geq \de$.
Using the notation $(z)^+$ for the representation of any $z\!\!\mod L$ within the interval $[0,L)$, the normalization equals
\begin{equation}\label{partition-fin}
Z(x) = 1 + \frac{N A}{L} - \frac{B}{L\, \ell_d}
\sum_{\al, \ga \neq \al}e^{-(x_\ga - x_\al)^+ / \ell_d}
\end{equation}
where we have introduced the form factors
\begin{eqnarray}
A &=& \int_{-\de}^\de \phi^- \phi^+(z)\,\id z \\&&
-\frac{1}{\ell_d} \int_{-\de}^\de \id z \int_{-\de}^z \id z'\,\phi^-(z) \phi^+(z')\,e^{(z'-z)/\ell_d}\nonumber\\
B &=& B^+ B^-, \quad \text{with}\;\;
B^\pm = \int_{-\de}^\de \phi^\pm(z)\,e^{z/\ell_d}\,\id z
\end{eqnarray}
that encode relevant details of the probe-colloid interaction; note their $\ve-$dependence is irrelevant for
\emph{locally} weakly driven colloids ($\ell_d \gg \de$).

For the stationary colloidal current we find
\begin{equation}\label{current}
j_x = \frac{\ve\rho^o}{\ze Z(x)}
\end{equation}
The bound $0 \leq j_x \leq \ve\rho^o / \ze$ yields $Z(x) \geq 1$, the equality holding true only for a homogeneous fluid without probes. Therefore the normalization quantifies the damping of the colloidal current by the probes: $j_x = j^{u\equiv 0} / Z(x)$.

\vspace{2mm}
\emph{Non-reactive forces ---}
The mean force on the
$\al$th probe as computed from~\eqref{forces} and \eqref{density-neq} is the sum of two generic components,
\begin{equation}\label{force-fin2}
f_\al(x) = f^\text{drift}(x) + f^\text{int}_\al(x)
\end{equation}
The first component is a ``drift'' force,
\begin{equation}\label{force-drift}
f^\text{drift}(x) = \ze A\, j_x 
\end{equation}
equal for all probes and proportional to the current.\\ 
The second component adds an effective probe force of order $\ve^2$,
\begin{equation}\label{force-int}
f^\text{int}_\al(x) = - \frac{\ze j_x B}{\ell_d}\,
\sum_{\ga \neq \al} e^{-(x_\ga - x_\al)^+/\ell_d}
\end{equation}
by which each probe is influenced
by all other probes which are ahead of it on the length scale $\ell_d$, but \emph{not vice versa}.
Therefore, this effective binary ``interaction'' is \emph{non-reactive} in the sense that it manifestly violates the action-reaction principle (Newton's Third Law). The absence of back-reaction clearly originates from the backward orientation of the density tails in the  driven colloidal fluid
(on the assumption $L \gg \ell_d$).

The effective non-reactive force can be both repulsive or attractive, depending on the sign of the form factor $B = B^+ B^-$. To obtain a repulsive force we need that both $B^\pm$ have the same sign, for which it is sufficient, e.g., that the probe-colloid interaction potential
is either completely positive ($u \geq 0$) or completely negative ($u \leq 0$). Another sufficient condition is that the temperature is large enough such that the two factors
$B^\pm = -2\be \int_0^\de u(z)\,\id z + O(\be^2)$ become nearly equal.
Attractive non-reactive forces ($B < 0$) can also be obtained, however they require a somewhat more subtle tuning of the model parameters.

We note that the non-reactivity of the effective interaction forces does not rely on the strong (global) driving condition
$L \gg \ell_d$. However, the latter condition of large separation of relevant length scales makes the non-reactivity effect fully developed, in the sense that the backward reaction entirely vanishes (beyond the microscopic scale $\de$).

\vspace{2mm}
\emph{Stability of crystal patterns ---}
Most interestingly, the driving-induced forces make equidistant arrangements of probes on the ring dynamically stable, provided the induced (non-reactive) probe-interaction is repulsive, $B>0$. This emergence of stability can be described within standard linear-stability theory:

Taking the probes as an overdamped dynamical system
$\Ga \dot x_\al = f_\al(x)$, $\al = 0,\ldots, N-1$, their equidistant configuration
\begin{equation}\label{equidistant}
x^*_\al(t) = v^* t + \frac{L}{N}\,\al\,\quad (\al\,\,\text{mod }N)
\end{equation}
forms a stationary cycle with steady rotation speed~\cite{SM}
\begin{equation}\label{velocity}
v^* = \frac{\ze j^*}{\Ga}\Bigl( A - \frac{B}{\ell_d} \frac{1}{e^{\frac{L}{N \ell_d}} - 1} \Bigr)
\end{equation}
proportional to the stationary current $j^* \equiv j_{x^*}$.
The linearized dynamics of the perturbation
$x_\al = x^*_\al + y_\al$ has the standard form
\begin{equation}\label{linearized}
\Ga \dot y_\al = \sum_\ga M_{\al\ga} y_\ga\,,\qquad
M_{\al\ga} = \frac{\partial f_\al(x^*)}{\partial x_\ga}
\end{equation}
The translational invariance of the composed probe-colloidal system reads
$f_\al(x + z) = f_\al(x)$
(with $z$ a simultaneous shift of all probes), whence the stiffness matrix satisfies
$M_{\al\al} = -\sum_{\ga\neq\al} M_{\al\ga}$. Moreover, since the probes are identical and equidistant, the off-diagonal elements ($\al \neq \ga$) are
$M_{\al\ga} = m_{\ga - \al}$ for
certain linear elasticity coefficients $m_\ga$. They play the role of effective ``spring constants'' though they are not symmetric $m_\ga \neq m_{-\ga}$,
due to the non-reactivity of the forces.

In the supplemental material \cite{SM} we derive a general form of the effective spring constants
which is valid for an \emph{arbitrary driving strength}, only requiring the probes to be separated. They are given by the exact formula
\begin{equation}\label{spring}
m_\al = \frac{\ze j^* B}{\ell_d^2\, (1 - e^{-L / \ell_d})}\,e^{-\frac{L}{N\ell_d} \al}\,,\quad
\al = 1,\ldots,N-1
\end{equation}
By translation invariance the linearized dynamics~\eqref{linearized} has a zero mode as it conserves 
$Y = \sum_\al y_\al$. 
We therefore speak about the stability of the configuration $y \equiv 0$ on the hypersurface $\sum_\al y_\al = 0$.  From \eqref{linearized} we check that 
$\La = \sum_\al y^2_\al$ is a Lyapunov function: For $B > 0$ all $m_\al>0$ in \eqref{spring} are positive and
$\dot\La(t) < 0$ unless $y \equiv 0$, proving stability of the crystal pattern. Analogously we prove instability for $B < 0$.

Interestingly,
both the current $j^*$ and the form factor $B$ are invariant under the transformation
$u \mapsto -u$ exchanging attractive for repulsive probe-colloid interactions, which then implies
$m^{-u}_\al = m^u_\al$. That is not true in general for the forces $f_\al(x)$, but the driving-reversal symmetry of the forces
$f^{-\ve}_\al(x) = f^\ve_\al(-x)$ implies the symmetry
$m^{-\ve}_\al = m^\ve_{N - \al}$.

For globally \emph{weakly} driven colloids ($L \ll \ell_d$) the exponential damping of $m_\al$ in $\alpha$ is negligible and the constants become almost homogeneous and of order $\ve^2$.
In contrast, for $L \gg \ell_d$ (globally \emph{strong} driving) we have
$m_\al = O(|\ve|^3)$ and the spring constants exhibit total asymmetry as
$m_{N-\al}$ becomes negligible for $1 \leq \al \ll N$.  
In general  $m_\al$ becomes negligible whenever
the distance $\al L/N$ between the probes $0$ and $\al$
is  large compared to the driving length scale $\ell_d = (\be\ve)^{-1}$. Therefore we need
$N \ell_d / L \gtrsim 1$ for the stabilization of the total crystal to be visible to be combined with the separation condition
$L/(2\de)  \geq N$.
That brings us in the intermediate regime of globally strong ($L\gg \ell_d$) yet locally not strong ($\ell_d\gtrsim 2\delta$) nonequilibrium alluded at before and where
the probe spacing $L/N$
does not exceed the dissipation length $\ell_d$.

Non-reactive interactions substantially modify the way the perturbed probes return to the crystal pattern. In the totally asymmetric case ($L \gg \ell_d$) the relaxation can be mapped exactly onto a compound Poisson process~\cite{SM}: Any local initial perturbation travels in the form of a Poisson wave  with a constant speed in the negative direction (this way creating an ``echo'' effect) and spreads in a diffusive manner, always keeping the zero mode fixed. The analysis~\cite{SM} shows a large time separation between the (fast) echo and the (slow) diffusion, the latter determining the relaxation time
\begin{equation}
\tau_\text{relax} = \frac{\Ga L^2}{\ze j^* B\, \Psi(\xi)}\,,\qquad \xi = \frac{L}{\ell_d N}
\end{equation}
with a dimensionless factor $\Psi(\xi)$ of order $O(1/\xi)$ for
\mbox{$\xi \ll 1$} (high number of probes per dissipation length), and of order $O(e^{-\xi})$ for $\xi \gg 1$. We observe that the relaxation time decreases with the number of probes, while it exponentially blows up whenever the probe spacing exceeds the dissipation length.

To estimate the steady fluctuations around the rigid crystal positions we suppose the friction and noise to be dominated by the interaction of the probes with the thermal environment at inverse temperature $\be$. That modifies \eqref{linearized} to
$\Ga \dot y_\al = \sum_\ga M_{\al\ga} y_\ga + (2\Ga/\be)^{1/2} \xi_{\al}$ with the $\xi_\al$'s independent standard white noise processes. A condition for the crystal to be thermally stable reads
$\langle (y_1 - y_0)^2 \rangle \ll (L/N)^2$. In the (arguably nearly optimal) regime
$L/N \approx \ell_d$ that requires~\cite{SM},
\begin{equation}\label{thermal}
\be\ze j^* B \gg 1
\end{equation}
which provides a lower bound on the colloidal current to generate a crystal pattern that is stable against thermal fluctuations.

\vspace{2mm}
\emph{Conclusions ---}
Exact solutions have been obtained for the induced forces on probes immersed in a colloidal medium at inverse temperature $\beta$ driven with force $\ve$ in a toroidal trap.   An important feature of the induced forces is that they violate the action-reaction principle
and their range coincides with the dissipation length
$\ell_d = (\beta\ve)^{-1}$. They can be either repulsive or attractive depending on the nature of the colloid-probe interaction.
The analysis shows an important interplay between the dissipation length $\ell_d = (\beta\ve)^{-1}$, the inverse probe density $L/N$, the probe-colloid interaction range $\delta$ and the system size $L$.
A crystal-like pattern of probes is dynamically stable when the probe-colloid interaction is either completely repulsive or completely attractive, in the regime of global nonequilibrium $L\gg \ell_d$ and where both the interaction range $\delta$ and the inverse probe density $L/N$ are not larger than the dissipation length $\ell_d$.  The action-reaction violation is manifested via a Poisson-wave return to the steady pattern after its perturbation.
The most stable patterns against thermal fluctuations are found in the regime
$L/N \approx \ell_d$ at large enough colloidal current.

\vspace{2mm}
\noindent {\bf Acknowledgments:}  
We thank Urna Basu for providing the figures and for many initial discussions.
KN acknowledges the support from the Grant Agency of the Czech Republic, grant no.~17-06716S.



\onecolumngrid

\newpage
\appendix

\section{Supplemental material}
\subsection{Stationary density of driven colloidal fluid}

An explicit solution to the stationary Smoluchowski equation~\eqref{stat}
can be obtained by direct integration
and it has the form of a marginal density
\begin{equation}\label{s-density}
\rho_x(\eta) = \int_0^L \mu_x(\eta,\eta')\,\id\eta'
\end{equation}
for the auxiliary two-colloid distribution
\begin{equation}\label{s-normalization}
\mu_x(\eta,\eta') = \frac{\rho^o L}{\Om(x)} e^{-\be W_x(\eta,\eta')}\,,\qquad
\Om(x) = \int_0^L \int_0^L
e^{-\be W_x(\eta,\eta')}\,\id\eta\,\id\eta'
\end{equation}
where
\begin{equation}
W_x(\eta,\eta') = U(x,\eta) - U(x,\eta') +
\begin{cases}
\ve (\eta' - \eta)^+ & \text{for } \ve \geq 0
\\
-\ve (\eta - \eta')^+ & \text{for } \ve \leq 0
\end{cases}
\end{equation}
Following the notation of the Letter, the shorthand $(\eta' - \eta)^+$ stands for
$\eta' - \eta\!\! \mod L \in [0,L)$.
The normalization factor $\Om(x)$ is related to the stationary current by
\begin{equation}\label{s-current}
j_x = \frac{\rho^o L (1 - e^{-\be\ve L})}{\be\ze\,\Om(x)}
\end{equation}
Another useful identity for the current is obtained by dividing equation~\eqref{stat} by
$\rho_x$ before integrating over the circle. We obtain
\begin{equation}\label{s-current-id}
j_x = \frac{\ve L}{\ze \int_0^L \frac{\id\eta}{\rho_x(\eta)}}
\end{equation}
which yields the bounds
\begin{equation}\label{s-current-bound}
0 < \frac{\ze j_x}{\ve\rho^o} \leq 1
\end{equation}
The upper bound obtained by convexity (the Jensen inequality)
turns into equality only if the colloidal density is constant,
which is verified only for $\partial U / \partial\eta = 0$ (i.e., no probes attached).\\

The total potential is
$U(x,\eta) = \sum_\al u(x_\al - \eta)$ with a generic probe-colloid interaction potential which is by assumption identical for all probes and satisfying
(i) $u(z) = 0$ whenever $|z| > \de$ (locality), and (ii) $u(-z) = u(z)$ (parity). We list relevant symmetries of the stationary colloidal medium:
\begin{enumerate}
	\item[(I)]
	translation invariance:
	$\rho_{x+z}(\eta+z) = \rho_x(\eta)$,
	$\Om(x+z) = \Om(x)$, and $j_{x+z} = j_x$ (with $(x+z)_\al = x_\al + z$ the simultaneous shift of all probes);
	\item[(II)]
	driving-reversal symmetry:
	$\rho^{-\ve}_x(\eta) = \rho^\ve_{-x}(-\eta)$;
	$\Om^{-\ve}(x) = e^{\be\ve L}\, \Om^\ve(-x)$, and $j^{-\ve}_x = -j^\ve_{-x}$;
	\item[(III)]
	$u-$reversal symmetry:
	$\Om^{-u}(x) = \Om^u(-x)$ and $j^{-u}_x = j^u_{-x}$
	(however the $u-$reversal of the density is nontrivial).
\end{enumerate}
In terms of the modified normalization function
\begin{equation}\label{s-partition-mod}
Z(x) = \frac{\be\ve}{L(1 - e^{-\be\ve L})}\,\Om(x)
\end{equation}
the current~\eqref{s-current} takes the form
\begin{equation}
j_x = \frac{\ve\rho^o}{\ze Z(x)}\,,\qquad
Z^{-\ve}(x) = Z^\ve(-x)
\end{equation}
and the upper bound in~\eqref{s-current-bound} reads
$Z(x) \geq 1$ with the equality if and only if $u \equiv 0$. Therefore $Z(x)$ is a damping factor for the colloidal current due to the interaction with the probes:
$j_x=j^u_x = j^{u \equiv 0} / Z(x)$.\\

Next we focus on the (globally) strong driving regime,
$L \gg \ell_d$ with
$\ell_d = (\be |\ve|)^{-1}$ the driving length scale, on the assumption that the probes are mutually separated
\begin{equation}
|x_\al - x_\ga| \geq 2\de \qquad \text{pairwise}
\end{equation}
Employing the Mayer expansion formulas
\begin{align}
e^{-\be U(x,\eta)} &= 1 + \sum_\al \phi^-(x_\al - \eta),\qquad
\phi^-(z) = e^{-\be u(z)} - 1
\\
\intertext{and}
e^{\be U(x,\eta)} &= 1 - \sum_\al \phi^+(x_\al - \eta),\qquad
\phi^+(x) = 1 - e^{\be u(z)}
\end{align}
the partition function~\eqref{s-partition-mod} for $\ve > 0$ reads, neglecting terms
$O(e^{-\be\ve L})$,
\begin{equation}
\nonumber
\begin{split}
Z(x) &= \frac{\be\ve}{L} \int_0^L \id\eta \int_{\eta}^{\eta + L}
\bigl[ 1 - \sum_\al \phi^+(x_\al - \eta) \bigr]\cdot \bigl[ 1 + \sum_\ga \phi^-(x_\ga - \eta') \bigr]\,
e^{-\be\ve(\eta' - \eta)}\,\id\eta'
\\
&= 1 + \frac{N}{L} \int_{-\de}^\de [\phi^- - \phi^+](z)\,\id z
-\frac{\be\ve}{L} \sum_{\al \ga} \int_{-\de}^{\de} \id z
\int_z^{z+L} \phi^+(z) \phi^-(z' + x_\al - x_\ga)\,
e^{-\be\ve(z' - z)}\,\id z'
\end{split}
\end{equation}
The second term in combination with the contributions $\al = \ga$ in the last term depend on the probes only via their density, whereas the contributions $\al \neq \ga$ add an exponentially damped dependence on the relative probe positions. Introducing the form factors
\begin{equation}
A = \int_{-\de}^\de \phi^- \phi^+(z)\,\id z -
\be\ve \int_{-\de}^\de \id z\, \phi^-(z)\,e^{-\be\ve z}\,
\int_{-\de}^z \phi^+(z')\,e^{\be\ve z} \id z'
\end{equation}
and
\begin{equation}
B = B^+ B^-\,,\qquad
B^\pm = \int_{-\de}^\de \phi^\pm(z)\,e^{\be\ve z}\,\id z
\end{equation}
the partition function takes the asymptotic form
\begin{equation}\label{s-partition-strong}
Z(x) = 1 + \frac{N A}{L} - \frac{\be\ve B}{L} \sum_{\al,\,\ga \neq \al} e^{-\be\ve (x_\ga - x_\al)^+}
\end{equation}
which is the formula~\eqref{partition-fin}.
Analogously we get, always up to $O(e^{-\be\ve L})$,
\begin{equation}
\begin{split}
\int_0^L e^{-\be W_x(\eta,\eta')}\,\id\eta' &=
e^{-\be U(x,\eta)} \int_\eta^{\eta + L} \bigl[ 1 - \sum_\al \phi^+(x_\al - \eta') \bigr]\,
e^{-\be\ve(\eta' - \eta)}\,\id\eta'
\\
&= e^{-\be U(x,\eta)} \bigl[ (\be\ve)^{-1} -
\sum_\al \Phi(x_\al - \eta)\,e^{-\be\ve(x_\al - \eta)} \bigr]
\end{split}
\end{equation}
where
\begin{equation}
\begin{split}
\Phi(z) = \int_{-\de}^z \phi^+(z')\,e^{\be\ve z'}\,\id z' =
\begin{cases}
B^+ & \text{for } z \geq \de
\\
0 & \text{for } z \leq -\de
\end{cases}
\end{split}
\end{equation}
is a smooth interpolation between zero and $B^+$ on the microscopic interval
$(-\de,\de)$. Combined with~\eqref{s-partition-strong} it yields the asymptotic formula~\eqref{density-neq} for the stationary colloidal density,
\begin{equation}\label{s-density-strong}
\rho_x(\eta) = \frac{\rho^o\,e^{-\be U(x,\eta)}}{Z(x)}\,
\bigl[1 - \be\ve \sum_\al \Phi(x_\al - \eta)\,e^{-\be\ve(x_\al - \eta)} \bigr]
\end{equation}

Note that in the globally strong but locally weak driving regime, $\be\ve\de \ll 1 \ll \be\ve L$, the $\ve-$dependence of the form factors is negligible and we can use simplified expressions
\begin{equation}
A_\ve = B^-_0 - B^+_0 + O(\be\ve\de),\qquad
B^\pm_\ve = B^\pm_0 + O\bigl((\be\ve\de)^2\bigr)\,,\qquad
B^\pm_0 = \int_{-\de}^\de \phi^\pm(z)\,\id z
\end{equation}

\subsection{\bf Driving-induced forces}

The transmission rule~\eqref{force-current} for the sum of all forces
\begin{equation}
\sum_\al f_\al(x) = -\sum_\al \int_0^L \frac{\partial U(x,\eta)}{\partial x_\al}\,\rho_x(\eta)\,\id\eta =
\int_0^L \frac{\partial U(x,\eta)}{\partial\eta}\,\rho_x(\eta)\,\id\eta
\end{equation}
immediately follows by integrating the Smoluchowski equation~\eqref{smoluchowski} over the ring.

To compute the individual force on the $\al$th probe in the strong-driving asymptotics we rewrite it as
\begin{equation}
\begin{split}
f_\al(x)
&= \frac{1}{\be} \int_0^L \rho_x(\eta)\,e^{\be U(x,\eta)}\,
\frac{\partial}{\partial x_\al}\bigl[ e^{-\be U(x,\eta)} \bigr]\, \id\eta
\\
&= -\frac{1}{\be} \int_0^L \rho_x(\eta)\,e^{\be U(x,\eta)}\,
\frac{\partial}{\partial\eta} \phi^-(x_\al - \eta)\, \id\eta
\\
&= \frac{1}{\be} \int_0^L \phi^-(x_\al - \eta)\,
\frac{\partial}{\partial\eta}\, \bigl[ \rho_x(\eta)\,e^{\be U(x,\eta)}\, \bigr]\,  \id\eta
\end{split}
\end{equation}
and by applying~\eqref{s-density-strong} we obtain, always up to $O(e^{-\be\ve L})$,
\begin{equation}
f_\al(x) =
\frac{\ve \rho^o}{Z(x)} \sum_\ga \int_0^L \phi^-(x_\al - \eta)\,
\bigl[\phi^+(x_\ga - \eta) - \be\ve\, \Phi(x_\ga - \eta)\,e^{-\be\ve(x_\ga - \eta)} \bigr]\,\id\eta
\end{equation}
The contribution $\ga = \al$ yields the $\alpha-$independent drift component
\begin{equation}\label{s-drift-strong}
f^\text{drift}(x) = \frac{\ve\rho^o A}{Z(x)} = \ze j_x A
\end{equation}
whereas the contributions $\ga \neq \al$ provide us with the interaction component
\begin{equation}
f^\text{int}_\al(x) = -\frac{\be \ve^2 \rho^o B}{Z(x)} \sum_{\ga \neq \al} e^{-\be\ve(x_\ga - x_\al)^+}
= -\frac{\ze j_x B}{\ell_d} \sum_{\ga \neq \al} e^{-(x_\ga - x_\al)^+ / \ell_d}
\end{equation}

In particular, for the equidistant configuration~\eqref{equidistant},
\begin{equation}
f^\text{int}(x^*) =
-\frac{\ze j_{x^*} B}{\ell_d} \frac{1}{e^{\frac{L}{N \ell_d}} - 1}
\end{equation}
which in combination with~\eqref{s-drift-strong} yields the total force
\begin{equation}
f(x^*) = \ze j_{x^*} \Bigl( A - \frac{B}{\ell_d} \frac{1}{e^{\frac{L}{N \ell_d}} - 1} \Bigr)
\end{equation}
on each probe, which then generates the steady velocity~\eqref{velocity} of the pattern rotation. An equivalent and more explicit expression is found directly from the transmission rule~\eqref{force-current} and formulas~\eqref{current} and \eqref{partition-fin}:
\begin{equation}
f(x^*) = \frac{\ve\rho^o L}{N} \Bigl[ 1 - \frac{1}{Z(x^*)} \Bigr]\,,\qquad
Z(x^*) = 1 + \frac{N A}{L} - \frac{\be\ve N B}{L} \frac{1}{e^{\frac{L}{N \ell_d}} - 1}
\end{equation}

\subsection{Stability of the crystal probe configuration}

In general the linear stiffness matrix reads
\begin{equation}\label{M-general}
\begin{split}
M_{\al\ga}(x) &= \frac{\partial f_\al(x)}{\partial x_\gamma}=
-\int_0^L \frac{\partial^2 U}{\partial x_\al \partial x_\ga}\,
\rho_x(\eta)\,\id\eta -
\int_0^L \frac{\partial U(x,\eta)}{\partial x_\al}\,
\frac{\partial\rho_x(\eta)}{\partial x_\ga}\,\id\eta
\\
&= \int_0^L \Bigl[ \beta\,u'(x_\al - \eta)\,u'(x_\ga - \eta) \,\rho_x(\eta)\,\id\eta
-u''(x_\al - \eta)\,\de_{\al\ga} \Bigr]
\\
&\phantom{***} - \beta \int_0^L
\int_0^L u'(x_\al - \eta)\,u'(x_\ga - \eta')\,
\mu_x(\eta,\eta')\,\id\eta\,\id\eta'
\\
&\phantom{***}+ f_\al(x)\,\frac{\partial{\log \Omega(x)}}{\partial x_\gamma}
\end{split}
\end{equation}
where $\mu_x(\eta,\eta')$ and $\Om(x)$ are the auxiliary density and its normalization~\eqref{s-normalization}, respectively.
For $\al \neq \ga$ and separated probes the first integral vanishes. By translational invariance
$\Om(x + z) = \Om(x)$ we have
\begin{equation}\label{Om_simple}
\sum_{\ga} \frac{\partial\log\Om(x)}{\partial x_\ga} = 0
\end{equation}
In particular, for the equidistant configuration of identical probes
$\partial\log\Om(x^*) / \partial x_\ga = 0$ for each probe, which makes vanish also the last term in~\eqref{M-general} and we are left with, for $\ve > 0$,
\begin{equation}
\begin{split}
M_{\al\ga}(x^*) &= -\frac{\be\rho^o L}{\Om^*}
\int_{0}^{L} \id\eta \int_{\eta}^{\eta + L} \id\eta'\,
u'(x^*_\al - \eta)\,u'(x^*_\ga - \eta')\,
e^{-\be U(x^*,\eta) + \be U(x^*,\eta') - \be\ve(\eta'-\eta)}
\\
&= -\frac{\be\rho^o L}{\Om^*}\,e^{-\be\ve(x^*_\ga - x^*_\al)^+}\,
\caI^-\, \caI^+
\end{split}
\end{equation}
where $\Omega^*=\Omega(x^*)$ and
\begin{equation}
\caI^- =
\int_{-\de}^{\de} \id z\,u'(z)\,e^{-\be u(z) + \be\ve z}
= \ve \int_{-\de}^{\de} \phi^-(z)\,e^{\be\ve z}\,\id z = \ve B^-
\end{equation}
and analogously
\begin{equation}
\caI^+ =
\int_{-\de}^{\de} \id\eta\,u'(z)\,e^{\be u(z) - \be\ve z}
= -\ve \int_{-\de}^{\de} \phi^+(z)\,e^{-\be\ve z}\,\id z = -\ve B^+
\end{equation}
Therefore the spring constants $m_\al = M_{0\hspace{0.5pt} \al}$, $\al = 1,2,\ldots,N-1$ are
\begin{equation}
m_\al = \frac{\be\ve^2\rho^o L\,B}{\Om(x^*)} e^{-\be\ve x^*_\al}
= \frac{\ga j^* B}{\ell_d^2 (1 - e^{-L / \ell_d})} e^{-\frac{L}{N \ell_d} \al}
\end{equation}
which is the formula~\eqref{spring}.\\

The linearized dynamical system
\begin{equation}
\Ga \dot y_\al = \sum_\ga M_{\al\ga} y_\ga =
\sum_{\ga = 1}^{N-1} m_\ga (y_{\al + \ga} - y_\al)
\end{equation}
conserves the zero mode $Y = \sum_\al y_\al$.  On the invariant hypersurface $\sum_\al y_\al = 0$ the configuration $y_\alpha\equiv 0$ is stable for $B > 0$ and is unstable for $B < 0$. To prove it formally, it suffices to check the Lyapunov property of the function
\begin{equation}
\La(y) = \frac{1}{2\Ga} \sum_\al y_\al^2
\end{equation}
Indeed,
\begin{equation}
\begin{split}
\dot \La(y) &= \sum_\al y_\al \dot y_\al
= \sum_{\ga=1}^{N-1} m_\ga \sum_{\al=0}^{N-1} y_\al (y_{\al+\ga} - y_\al)
\\
&= -\frac{1}{2} \sum_{\ga=1}^{N-1} m_\ga \sum_{\al=0}^{N-1} (y_{\al+\ga} - y_\al)^2
\end{split}
\end{equation}
and its sign coincides with $-\text{sgn}\, B$.
Note this is equivalent to positivity (or negativity) of the symmetric part of the stiffness matrix
$M + M^T \geq 0$, which is a sufficient condition for stability; see e.g.~\cite{khalil}.

\subsection{Relaxation to the crystal pattern}
\label{formation}

It remains to be understood \emph{how} an initial perturbation relaxes to the equidistant stationary configuration, i.e., what is the relaxation profile solving the linear equations~\eqref{linearized}. Without restriction it suffices to consider the initial condition
$y_0(0) = \kappa$ and $y_\al(0) = 0$, $\al \neq 0$ for some small distance $\kappa$. As a variant to standard analytical methods, we establish all relevant features of the relaxation profile by a simple trick as now follows.

To simplify the notation, in the sequel we write the spring constants as
$m_\al = D\, e^{-\xi \al}$, with the shorthand
\begin{equation}\label{D-al}
D = \frac{\be^2 \ve^3 \rho^o B}{Z^*}\,,\qquad \xi = \frac{\be\ve L}{N}
\end{equation}
and $Z^*=Z(x^*)$.  Let us consider the \emph{compound Poisson process}
$\caX$ on $\bbN = \{0,1,2,\ldots\}$ started from
$\caX(0) = 0$, with transition rates
$\la(\al \rightarrow \al + \ga) = m_\ga/\Gamma$ for all $\al \geq 0$, $\ga>0$, and zero otherwise. The time-dependent probabilities
$p_\al(t) = \bsP[\caX(t) = \al]$ satisfy the Master equation
(with $p_\al \equiv 0$ whenever $\al<0$)
\begin{equation}
\begin{split}
\dot p_\al &= \sum_{\ga>0} \bigl[\, p_{\al-\ga}\,\la(\al-\ga \rightarrow \ga) -
p_\al\,\la(\al \rightarrow \al+\ga) \,\bigr]
\\
&= \frac{1}{\Ga}\,\sum_{\ga>0} m_\ga\,[\,p_{\al-\ga} - p_\al\,]
\end{split}
\end{equation}
to be solved with the initial condition
$p_\al(0) = \de_{\al 0}$. By replacing $\al$ with $-\al$ it gets the form
\begin{equation}
\dot p_{-\al} = \frac{1}{\Ga} \sum_{\ga>0} m_\ga\,[\,p_{-(\al+\ga)} - p_{-\al}\,]\,,\qquad
\al=0,-1,-2,\ldots
\end{equation}
Comparing with~\eqref{linearized} we observe that the
$y_\al(t)$, $\al = 0,-1,-2,\ldots (\hspace{-2.5mm}\mod N)$, solve a similar problem on the circle (for convenience now run in the opposite direction). More precisely we have exactly that
\begin{eqnarray}\label{poisson-rel}
y_{-\al}(t) &=& \kappa\,\sum_{\al=0}^{+\infty} p_{\al + kN}(t)\,,\qquad
\al = 0,1,\ldots,N-1\nonumber\\
&=& \kappa\,\bsP[\caX(t) = \al\;\, \text{mod} \, N]
\end{eqnarray}
We can thus obtain the solution to~\eqref{linearized} via solving the compound Poisson process
$\caX(t)$, which is well-known --- see e.g.~\cite{cp}. The latter is characterized by a Poisson counter with intensity
\begin{equation}
\La = \frac{1}{\Ga}\,\sum_{\al>0} m_\al = \frac{D}{\Ga} \frac{1}{e^\xi - 1}
\end{equation}
and by geometrically distributed jumps $\caJ$, with probabilities
$\bsP[\caJ = \al] = (e^\xi - 1)\,e^{-\xi \al}$ and with the first two moments
\begin{equation}
\langle \caJ \rangle = \frac{1}{1 - e^{-\xi}}\,,\qquad
\langle \caJ^2 \rangle = \frac{1 + e^{-\xi}}{(1 - e^{-\xi})^2}
\end{equation}
Then by applying standard results on Poisson processes,~\cite{cp}, we immediately obtain
\begin{equation}
\begin{split}
p_0(t) &= e^{-\La t} = e^{-\frac{D t}{e^\xi - 1}}
\\
\langle \caX(t) \rangle &= \langle \caJ \rangle\,\La t = \frac{e^\xi}{(e^\xi - 1)^2}\,
\frac{D t}{\Ga}
\\
\text{Var}\,\caX(t) &= \langle \caJ^2 \rangle\,\La t = \frac{e^\xi (e^\xi + 1)}{(e^\xi - 1)^3}\,
\frac{D t}{\Ga}
\end{split}
\end{equation}
In words, the process $\caX(t)$ leaves the origin and on average moves with speed
$\langle \caJ \rangle\,\La$ while diffusing. Therefore by the correspondence~\eqref{poisson-rel}, the initial perturbation $y$ located at the first probe starts to decay exponentially exponentially fast and moves in the negative direction (with respect to the drift).  It reaches again the origin in a somewhat delocalized form, and so on until it spreads over the whole circle. This way we detect three \emph{a priori} different times scales:\\
(I) the characteristic time of initial decay of the perturbation at the origin,
\begin{equation}
\tau_0 = \frac{1}{\La} = \frac{\Ga}{D} (e^\xi - 1)
\end{equation}
(II) the period of the \emph{echo effect} after which the Poisson wave visits the origin,
\begin{equation}
\tau_\text{echo} = \frac{N}{\langle \caJ \rangle \La} =
\frac{(e^\xi - 1)^2}{e^\xi}\,\frac{\Ga N}{D}
\end{equation}
(III) the relaxation time defined as a characteristic time in which the wave spreads over the whole circle,
\begin{equation}\label{t-relax}
\tau_\text{relax} = \frac{N^2}{\langle \caJ^2 \rangle\,\La} =
\frac{(e^\xi - 1)^3}{e^\xi (e^\xi + 1)}\,\frac{\Ga N^2}{D}
\end{equation}
Observing that
\begin{equation}
\frac{\tau_\text{echo}}{\tau_0} = (1 - e^{-\xi})\,N\,, \qquad
\frac{\tau_\text{relax}}{\tau_\text{echo}} = \frac{e^\xi - 1}{e^\xi + 1}\,N
\end{equation}
we see that all three time scales are well separated (note that
$\be\ve L = \xi N \gg 1$ by assumption) unless $N$ is small.
Whenever $N \gg 1$ we always have
\begin{equation}
\tau_0 \ll \tau_\text{echo} \ll \tau_\text{relax}
\end{equation}
showing that we have indeed localized distinct time scales which are all relevant in the description of the relaxation of a local perturbation to the crystal patterns.

\subsection{Thermally induced fluctuations}

To estimate the role of thermal fluctuations we study stationary solutions of the noisy linear dynamics
\begin{equation}\label{nodb}
\Ga \dot y_\al = \sum_\ga M_{\al\ga} y_\ga + \Bigl( \frac{2\Ga}{\be} \Bigr)^\frac{1}{2} \xi_\al
\end{equation}
By translation invariance the stiffness matrix $M$ is diagonalizable via a unitary matrix and therefore it is a normal matrix. As a consequence, the stationary distribution coincides with the Boltzmann distribution,
\begin{equation}
\nu(y) = \frac{1}{\caZ}\,e^{-\be V(y)}
\end{equation}
for the effective potential
\begin{equation}
V(y) = \frac{1}{2} \sum_{\al,\ga} y_\al M_{\al\ga} y_\ga =
\frac{1}{4} \sum_{\ga > 0} m_\ga \sum_\al (y_{\al + \ga} - y_\al)^2
\end{equation}
despite the absence of detailed balance in \eqref{nodb}; see \cite{njp1} for a similar scenario.  Therefore, 
\begin{equation}
\frac{1}{2} \sum_{\ga > 0} m_\ga \bigl\langle (y_\ga - y_0)^2 \bigr\rangle  = \frac{1}{\be} 
\end{equation}
Assuming for simplicity that $m_\ga$ is negligible for $\ga \geq 2$, we get
\begin{equation}\label{variance}
\bigl\langle (y_1 - y_0)^2 \bigr\rangle \approx \frac{2}{\be m_1} =
\frac{2\ell_d^2}{\be\ze j^* B} e^\xi\,,\qquad \xi = \frac{\be\ve L}{N} \gtrsim 1
\end{equation}
For $\xi \approx 1$, i.e. $L/N \approx \ell_d$, the condition of thermal stability
$\langle (y_1 - y_0)^2 \rangle \ll (L/N)^2$ yields the inequality~\eqref{thermal}. Obviously, for
$L/N \gg \ell_d$ the thermal fluctuations exponentially blow up along with an exponential growth of the relaxation time. On the other hand, for
$L/N \ll \ell_d$ the variance~\eqref{variance} goes as $O(N^{-1})$ which further strengthens the condition~\eqref{thermal}, showing that the regime
$L/N \approx \ell_d$ is nearly optimal.


\end{document}